\documentclass[aps,preprint]{revtex4}
\usepackage{graphicx}
\begin{document}

\title{Simple models of small world networks with directed links}

\author{A. Ramezanpour}
\email{ramzanpour@mehr.sharif.edu}
 \affiliation{Department of
Physics, Sharif University of Technology,P.O.Box
11365-9161,Tehran,Iran}

\author{V. Karimipour }
\email{vahid@sharif.edu}
 \affiliation{Department of Physics,
Sharif University of Technology,P.O.Box 11365-9161,Tehran,Iran}

\date{\today}

\begin{abstract}
We investigate the effect of directed short and long range
connections in a simple model of small world network. Our model
is such that we can determine many quantities of interest by an
exact analytical method. We calculate the function $V(T)$,
defined as the number of sites affected up to time $T$ when a
naive spreading process starts in the network. As opposed to
shortcuts, the presence of un-favorable bonds has a negative
effect on this quantity. Hence the spreading process may not be
able to affect all the network. We define and calculate a
quantity named the average size of accessible world in our model.
The interplay of shortcuts, and un-favorable bonds on the small
world properties is studied.
\end{abstract}

 \maketitle

\section{Introduction}
\begin{figure}
\includegraphics[width=8cm]{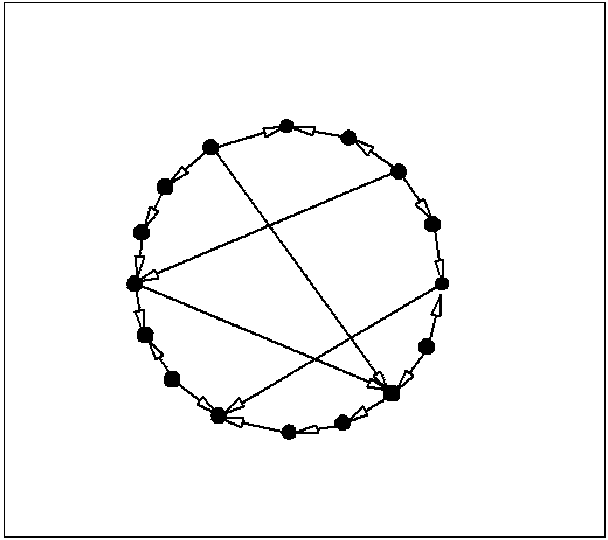}
    \caption{A WS-type model of directed network.}\label{fig-1}
\end{figure}

Real life networks, whether made by nature, (e.g. neural,
metabolic and ecological networks) \cite{fhcs,bm,jtao,kas,plh},
or made by human (e.g. the World Wide Web, power grids, transport
networks and social networks of relations between individuals or
institutes) \cite{ajb,n2,deb}, have special features which is a
blend of those of regular networks on the one hand and completely
random ones on the other hand. To study any process in these
networks,(the spreading of an epidemic in human society, a virus
in the internet, or an electrical power failure in a large city,
to name only a few), an understanding of their topological and
connectivity properties is essential (for a review see \cite{ab1}
and references therein). Recently obtained data from many real
networks show that like random networks \cite{b,asbs}, they have
low diameter, and like regular networks, they have high
clustering. Since the pioneering work of Watts and Strogatz
\cite{ws}, these networks have attracted a lot of attentions  and
have been
studied from various directions \cite{bw,nmw,ka,z,jb}.\\
In contrast to most of the models studied so far, many real
networks like the World Wide Web, neural, power grids, metabolic
and ecological networks have directed one-way links
\cite{bkmr,t2,ste,jtao}. These types of networks may have
significant differences in both their static and dynamic
properties with the Watts-Strogatz (WS) model and its variations
\cite{slr,t1,t2}. The presence of directed links affects strongly
many of the properties of a network. For example, for the same
pattern of shortcuts, the average shortest path in an directed
network is longer than that in an undirected one, due to the
presence of bonds with the wrong directions (blocks) in many
paths. So is
the spreading time of any dynamic effect on the lattice.\\
Consider the quantity $V(T)$ defined as the average number of
sites which are visited at least once when we start a naive
spreading process at a site and continue it for $T$ steps. Note
that we mean an average over an ensemble of networks and initially
infected sites and by the naive spreading process we mean that at
each step of the spreading process all the neighbors of an
infected site are equally infected. The quantity $V(T)$ may be
taken as a crude approximation for the number of people who have
been infected by a contiguous disease after $T$ time steps has
elapsed since the first person has been infected. Clearly this is
a simplification of the real phenomena, since in real world a
disease may not affect an immunized person or may not transmit
with certainty in a contact. However as a first approximation,
$V(T)$ may give a sensible measure of the effect in the whole
network. Since in an directed network, an effect only spreads to
those neighbors into which there are correctly directed links,
there will be pronounced differences in this important quantity
between an directed and an undirected network. As a concrete
example consider a ring with $N$ sites, without any shortcuts,
where to emphasize the absence of shortcuts, we denote $V(T)$ by $
V_0(T)$. If all the links have the same direction, we have $
V_0(T) = T$, and if all of them are bidirectional, we have $V_0(T)
= 2 T $. In both cases the whole lattice gets infected after a
finite time. However if the links are randomly directed then
$V_0(T)$ may be much lower and furthermore, there is a finite
probability that
only a small fraction of the whole lattice gets infected.\\
Adding shortcuts to this ring of course has a positive effect on
the spreading. In a sense we have a chance to see the interplay
of two different concepts of small worlds in these networks. The
size of the world as a whole may be small due to the ease of
communication with the remote points provided by long range
connections, however the world accessible to an individual may be
small due to the absence of properly directed links to connect it
to the outside world.\\
It is therefore natural to ask how the presence of directed links
and (or) directed shortcuts affects quantitatively the small world
properties of a network? How we can make a simple model of a small
world network with such random directions? A WS-type model for
these networks may be as shown in figure (\ref{fig-1}). However
due to their complexity, these networks should usually be studied
by numerical or simulation methods and they seldom amend
themselves to exact analytical treatment.

\subsection{The aim, structure and results of this paper}
\begin{figure}
\includegraphics[width=8cm]{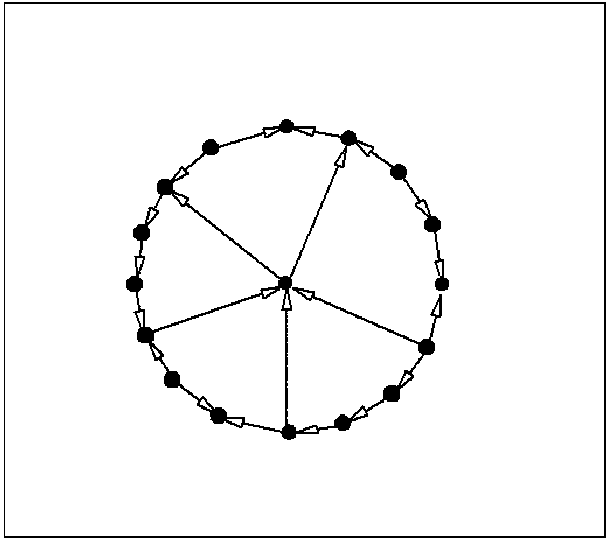}
    \caption{A simple substitute for the network of figure (\ref{fig-1}). }\label{fig-2}
\end{figure}

\begin{figure}
\includegraphics[width=8cm]{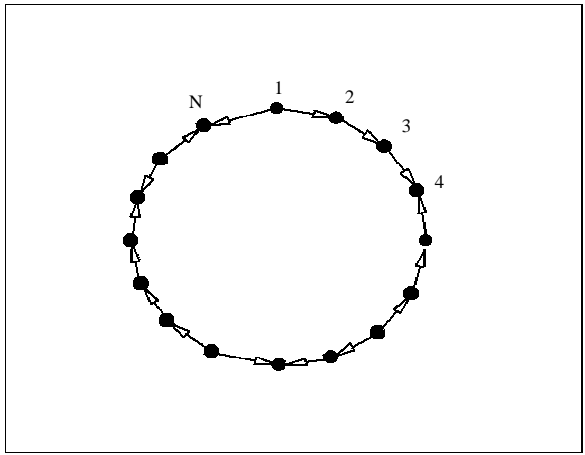}
    \caption{A regular ring with randomly directed links, without shorcuts.
    You can see also the accessible world of site $1$. }\label{fig-3}
\end{figure}

As we will show in this paper, with slight simplification one can
introduce simpler models which although retain most of the small
world features, are still amenable to analytical treatment. This
is what we are trying to do in this paper. In this paper we
introduce one such model following our earlier work \cite{kr}
which was in turn inspired by the work of \cite{dm}.  The basic
simplifying feature of these networks is that all the shortcuts
are made via a central site, figure (\ref{fig-2}). For such a
network many of the small world quantities, can be calculated
exactly.
In particular, once $V(T)$ defined above is calculated, many other
quantities like the average shortest path between two sites can be
obtained. An exact calculation of $V(T)$ is however difficult for
the case where both the shortcuts and the links have random
directions. We therefore proceed in two steps to separate the
effects of randomness in the two types of connections. First, in
section 2, we remove the shortcuts and calculate exactly $V(T)$
for a ring with random links, figure (\ref{fig-3}).
 To emphasize
the absence of shortcuts we denote this quantity by $ V_0(T)$.
Note that $V_0(T)$ depends only on the structure of the underlying
ring and its short-range connections. Then in section 3, we
consider only the effects of randomly directed shortcuts, that is
we let directions of the links on the ring to be regular and fixed
say clockwise, and calculate exactly $V(T)$, where again for
emphasis on the shortcuts we denote this quantity by $S(T)$. \\
We then argue, in section 4 that in the scaling limit where the
number of sites goes to infinity with the number of shortcuts kept
finite, most of the spreading takes place via the links and only
from time to time it propagates to remote points via the
shortcuts. In this limit it is plausible to suggest a form for
$V(T)$ which takes into account the effect of both the random
links and the shortcuts in the form $ V(T) = S(V_0(T))$. This may
not be an exact relation but as we will see it will give a fairly
good approximation of $V(T)$, as shown by the agreement of our
analytical results and the results of simulations. This then means
that in more complicated networks, one can separate the effects of
short and long range connections and superimpose their effect in a
suitable way. We conclude the paper with a discussion.

\section{Exact calculation of $V_0(T)$ in a ring with random bonds}
Consider a regular ring of $N$ sites whose bonds are directed
randomly. Each link may be directed clockwise with probability
$r$, counterclockwise with probability $\ell$, and bidirectional
with probability $ 1-r-\ell$.\\ Thus we have a problem similar to
bond percolation in a small world network.
 Suppose that at time $T=0$ site number $1$ is
infected with a virus. We ask the following questions: \\
After $T$ seconds how many sites have been infected on the
average? What is the average speed of propagation of this decease
in the network? These questions have obvious answers for rings
with regularly directed or bi-directional bonds, namely the number
of infected sites are respectively $ T$ and $2T$, with
corresponding speeds of propagation being $ 1$ and $2$. In the
randomly directed network, the situation is different. For example
if both neighbors of site $1$ are directed into this site, this
site can not affect any other site of the network. Such a site
being effectively isolated has an {\it {accessible world}} \cite{bkmr} of zero
size ( figure\ref{fig-3}). To proceed with exact calculation,
consider the right hand side of site $1$. The probability that
exactly $k < T $ extra sites to the right have been infected is $
P_+(k):= (1-\ell)^k \ell $, and the probability that exactly $ T $
extra sites have been infected is $ P_+(T):= (1-\ell)^T$.
Therefore the average number of extra sites infected to the right
of the original site is
\begin{eqnarray}\label{1}
  V^+_0(T) &=& \sum_{k=1}^T kP_+(k)= T(1-\ell)^T + \sum_{k=0}^{T-1} k(1-\ell)^k\ell
  \cr
  &=& T(1-\ell)^T + \frac{1}{\ell}[ 1-\ell +(1-\ell)^T(\ell-1-\ell
  T)].
\end{eqnarray}
Going to the large $N$ limit where,
\begin{equation}\label{scaling}
N\rightarrow \infty, \ \ \ell\rightarrow 0  \ \ \  \mu := \ell N,
\ \ \  t:= \frac{T}{N}, \ \ \  \upsilon_+(t):= \frac{V^+_0(T)}{N}
\end{equation}
we find the simple result
\begin{equation}\label{d}
  \upsilon^+_{0}(t) = \frac{1}{\mu}(1-e^{-\mu t})
\end{equation}
The same type of reasoning gives the number of sites infected to
the left $ \upsilon^-_0(t)$ and thus the total number of infected
sites will be:
\begin{equation}\label{dd}
  \upsilon_{0}(t) = \frac{1}{\mu}(1-e^{-\mu t})+ \frac{1}{\lambda}(1-e^{-\lambda t})
\end{equation}
where $\lambda := r N $. What are the meaning of the scaled
variables? The parameter $ \mu$ is the total number of sparse
blocked sites in the way of propagation to the right, with a
similar meaning for $ \lambda$. $ \upsilon_{0} (t) $ is the fraction
of infected sites up to time $ t $. In a bidirectional lattice,
all the sites could be infected after the passage of $ T =
\frac{N}{2} $ seconds, or at $ t = \frac{1}{2}$ and if $t$ passes
$ \frac{1}{2}$, some of the sites become doubly visited.
Therefore it is plausible for the sake of comparison to define a
quantity in our ring, namely the average size of the
{\it{accessible}} world as $ \upsilon_{0}^{\it{acc}}:=
\upsilon_{0}(\frac{1}{2})$, which turns out to be:
\begin{equation}\label{ddd}
  \upsilon_{0}^{\it{acc}} = \frac{1}{\mu}(1-e^{-\frac{\mu}{2}})+
  \frac{1}{\lambda}(1-e^{-\frac{\lambda}{2}})
\end{equation}
It is seen that the presence of only a small number of blocked
bonds causes a significant drop in the average size of this
accessible world. For example a value of $ \lambda = \mu = 4 $
leads to $\upsilon_{0}^{\it{acc}}\sim 0.4 $. The long range
connections (shortcuts) make the world small with the ease of
communication they provide, however blockades make the world
small in this new sense.\\
The speed of propagation is found from
\begin{equation}\label{dddd}
  \dot{\upsilon}_{0}(t) = e^{-\mu t}+
  e^{-\lambda t}.
\end{equation}
In the symmetric case where $ \lambda = \mu $, equation
(\ref{dd}) simplifies to:
\begin{equation}\label{dds}
  \upsilon_{0}(t) = \frac{2}{\mu}(1-e^{-\mu t})
\end{equation}
with
\begin{equation}\label{dddds}
\dot{\upsilon}_{0}(t) = 2 e^{-\mu t}
\end{equation}
Note that at the early stages of spreading when $ \mu t << 1$,
and the effects of blocked bonds has not yet been experienced,
the infection propagates with speed equal to $2$ as in a regular
network. The effect of blocking comes into play when $ t $
becomes comparable to $\frac{1}{\mu} $.\\
As a few number of shortcuts may enhance the speed of
propagation, a few number of blocked bonds may have the opposing
effect. First the blocks reduce the speed of propagation as is
clear from (\ref{dddd}) and second and more importantly they
reduce the number of accessible sites, or the size of accessible
world. It will thus be of interest to see how these two effects
compete in a random network where there are both shortcuts and
blocks. We will study this in the final section of this paper. To
this end we first study the effect of directed shortcuts in an
otherwise regular ring with no blocks.

\section{The long range connections}
\begin{figure}
\includegraphics[width=8cm]{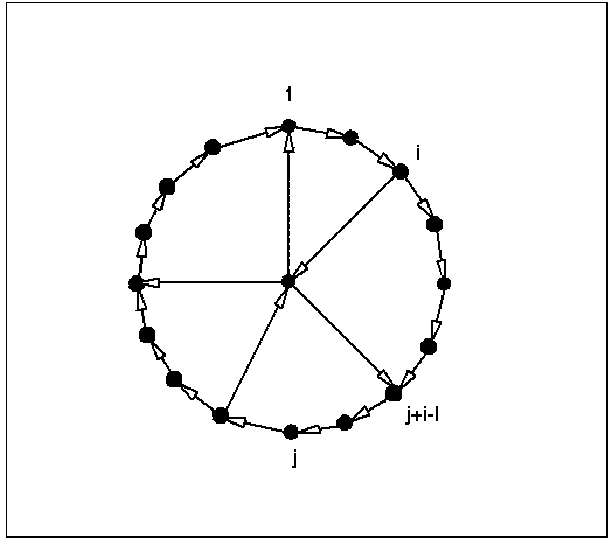}
    \caption{ Randomly directed shortcuts added to a ring with clockwise links}\label{fig-4}
\end{figure}

\begin{figure}
\includegraphics[width=8cm]{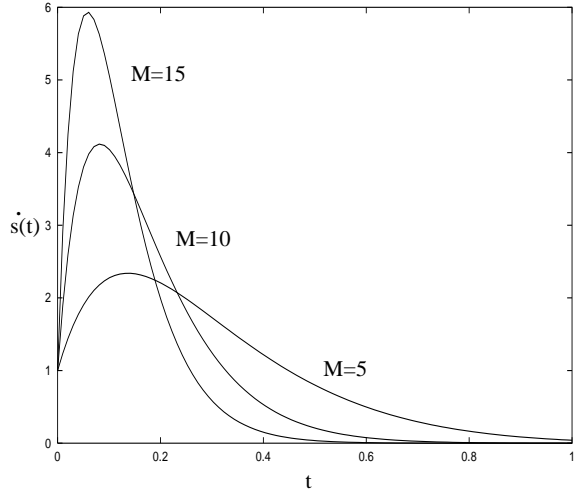}
    \caption{The speed of propagation in a ring, for
several values of randomly directed shortcuts.}\label{fig-5}
\end{figure}

In this section we are to consider only the effect of randomly
directed shortcuts in the spreading process and obtain exactly
the function $ S(T)$ for this network, figure (\ref{fig-4}).
Note that this function has the same meaning as $V(T)$, except
that for emphasis on the role of shortcuts in it we have adopted a
new name for it. We fix a regular clockwise ring. Between a site
and the center there is a shortcut going into the center with
probability $p$ and out of the center with probability $q$. The
site remains unconnected to the center with probability $1-p-q$.
The average number of connections into and out of the center are
respectively $M_i:=Np$
and $M_o:=Nq$.\\
Consider sites $1$ and $j$. We want to find the probability that
the shortest path between these two sites be of length $l$, a
probability which we denote by $ P(1,j;l)$. A typical shortest
path of length $l$ connecting these two nodes is shown in figure
(\ref{fig-4}), where the first inward connection to the center
occurs at site $i$ and the last outward connection from the
center occurs at site $j+i-l$. Such a path occurs with probability
$(1-p)^{i-1}pq(1-q)^{l-i}$. Summing over all such configurations
gives us the probability for the shortest path between sites $1$
and $j$
to be of length $l$. For $l \neq j-1$, we have:\\
\begin{equation}\label{16}
p(1,j;l\neq
j-1)=\sum_{i=1}^{l}(1-p)^{i-1}pq(1-q)^{l-i}=pq[\frac{(1-p)^l}{q-p}+\frac{(1-q)^l}{p-q}],
\end{equation}
and $p(1,j;j-1)$ is determined from normalization:
\begin{equation}\label{normalization2}
  P(1,j;j-1) = 1-\sum_{l=1}^{j-2} P(1,j;l)=
  \frac{1}{p-q}\Big(p(1-q)^{j-1}-q(1-p)^{j-1}\Big)
\end{equation}
Note that $p(1,j;l\ne j-1)$ dose not depend on $j$, a property
which is true for standard small world networks \cite{aks}.\\
Now consider a naive spreading process starting at site $1$. The
number of sites affected up to time $ T$, denoted by $ S(T)$,
builds up in two ways, via the links on the ring and via the
shortcuts. The first way gives a contribution $T+1$ and the second
way gives a contribution $(N-T-1)\sum_{l=1}^{T} p (1,j;l)$
\cite{aks} where $(N-T-1)$ is the number of sites beyond direct
reach at time $T$ which has been multiplied by the probability of
any of these sites being at a distance shorter than $T$ to site
$1$ via a shortcut. Putting
this together we find:\\
\begin{eqnarray}\label{17}
S(T)= T+1+(N-T-1)\sum_{l=1}^{T} P(1,j;l) \\ \nonumber = N+(N-T-1)
[\frac{q}{p-q}(1-p)^{T+1}+ \frac{p}{q-p}(1-q)^{T+1}]
\end{eqnarray}
In the scaling limit where $N\rightarrow \infty, \ \ p, q
\rightarrow 0, \ \ $ where  $M_i$ and $ M_o $ are kept fixed and $
s(t) := \frac{S(T)}{N}$, we find:
\begin{equation}\label{18}
s(t)=1-\frac{1-t}{M_i - M_o}\Big(M_ie^{-M_o t}- M_o e^{-M_it}\Big)
\end{equation}
In the symmetric case where $ M_i = M_o = M $ this  equation
simplifies to:
\begin{equation}\label{18s}
s(t)=1-(1-t)(1+Mt)e^{-Mt}
\end{equation}
with the speed of propagation
\begin{equation}\label{speed2s}
\dot{s}(t) = e^{-Mt} (1+ Mt + M^2t-M^2t^2)
\end{equation}
Figure (\ref{fig-5}) shows the speed of propagation as a function
of time for several values of $M$.

\section{The spreading effect in a directed small world network}
\begin{figure}
\includegraphics[width=8cm]{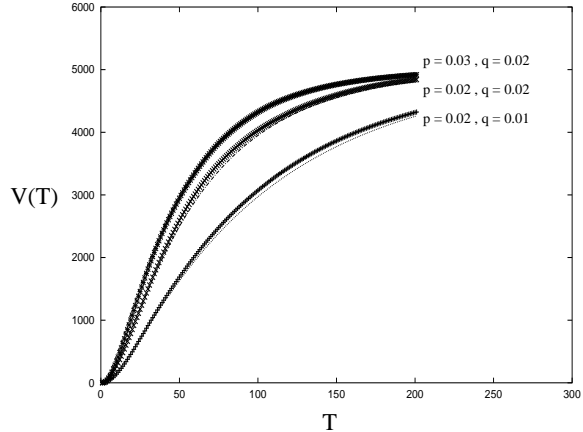}
\caption{$V(T)$ for a fully random network in the case $N=5000,
r=0.02,l=0$. Analytic results(lines) versus simulations(symbols)
which have been averaged over $1000$ realizations of the
network.}\label{fig-6}
\end{figure}

\begin{figure}
\includegraphics[width=8cm]{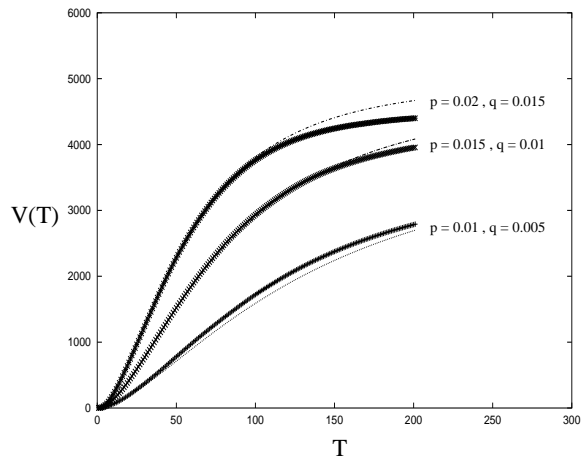}
 \caption{$V(T)$ for a fully random network in the case
$N=5000, r=l=0.005$. Analytic results(lines) versus
simulations(symbols) .}\label{fig-7}
\end{figure}

We now come to the problem of composing both the blocks and the
shortcuts in a model of small world network. That is we consider
the ring of figure (\ref{fig-2}) where randomly directed shortcuts
are added to a ring with randomly directed links. We can not
obtain exact expressions for this network from first principle
probability considerations. However we can obtain expressions for
$ \upsilon(t)$ in the scaling limit by a heuristic argument and
compare our results with those of simulations. Consider equation
(\ref{18s}). This equation shows how the presence of $2M$ randomly
directed shortcuts in a regular clockwise ring affects the
spreading effect. On the other hand we know that the number of
sites infected up to time $t$ in the absence of shortcuts, has
changed to $ \upsilon_0(t)$. Due to the rarity of shortcuts
compared to the regular bonds, most of the spreading takes place
via the local bonds, the role of shortcuts is just to make
multiple spreading processes happen in different regions of the
network. This role is the same whatever the underlying lattice
is, and therefore for a general network, at least in the scaling
regime, we can assume that equation (\ref{18s}) can be elevated
to $\upsilon(t) = s(\upsilon_0(t))$, i.e;
\begin{equation}\label{eee}
  \upsilon(t) = 1-
  (1-\upsilon_0(t))(1+M\upsilon_0(t))e^{-M\upsilon_0(t)}.
\end{equation}
For a fully random network where $2M$ randomly directed shortcuts
are distributed on a ring with already random links, we assume
that this relation holds true with $\upsilon_0(t)$ taken from
(\ref{dd}). This suggestion may not provide an exact solution
for the network, however we think it provides a fairly good
approximation. In fact exact solution for the case where all the
links on the ring are bidirectional is possible and it confirms
the above ansatz, that is we obtain an exact expression only by
setting $ \upsilon_0(t) = 2t $ in the above formula. Moreover
this separation of the effect of short and long range connections
may be also useful in more complicated networks. Whether this
assumption is plausible or not can be checked by comparison with
simulations. The results of simulations are compared with those
of equations (\ref{dd}) and(\ref{18}) in figure (\ref{fig-6}) and
(\ref{fig-7}).

\section{Static properties}
\begin{figure}
\includegraphics[width=8cm]{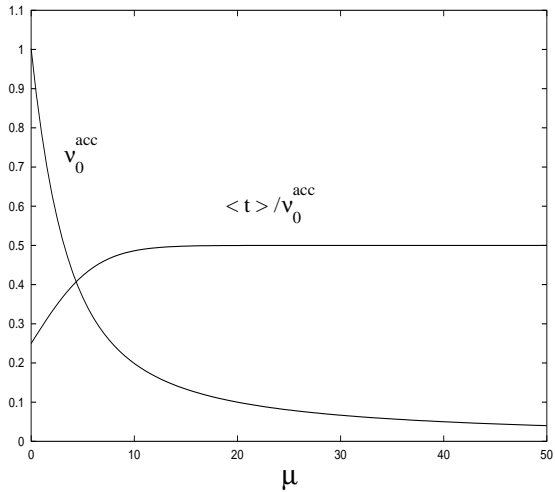}
\caption{ The average size of accessible world and the average
shortest path for a regular ring with randomly directed bonds
without shortcuts.}\label{fig-8}
\end{figure}

\begin{figure}
\includegraphics[width=8cm]{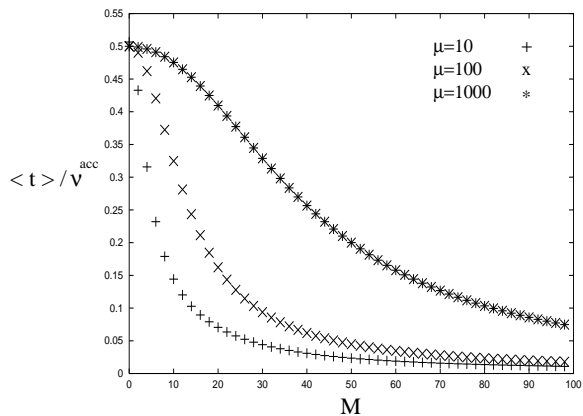}
\caption{The average shortest path for a fully random network.
}\label{fig-9}
\end{figure}

Once the functions $ V(T)$ or $ \upsilon(t)$ are obtained, the
static properties of the network i.e., the average shortest path
between two arbitrary sites and its probability distribution can
be calculated directly.\\ Since  $V(T)$ by definition is the
number of sites whose shortest distance to site $1$ is less than
or equal to $T$, we find the number of sites whose shortest
distance is exactly $T$ to be $ V(T)-V(T-1)$. Since site $1$ is
an arbitrary site, we find the probability distribution of the
shortest distance between two arbitrary sites which are
accessible to each other as: $ P(T)=
\frac{V(T)-V(T-1)}{V_{{\it{acc}}}} $, where $ V_{{\it{acc}}} $ is
the average
size of the accessible world.
(There is of course a slight approximation here in that we are
taking averages of the denominator and numerator separately.)\\
For a regular ring with shortcuts, $ V_{{\it{acc}}} = N $, since
all the sites are accessible. We will discuss the case of random
rings in the sequel. In the scaling regime the above formulas
transform to:
\begin{equation}\label{static1}
  P(t)=\dot{\upsilon}(t).
\end{equation}
Note that $P(t)$ is normalized, i.e. $ \int_{0}^{1} P(t)dt =
\upsilon(1)-\upsilon(0) = 1 $. The average shortest path for the
network of figure (\ref{fig-4}) when $M_i=M_o=M$, turns out to be:
\begin{equation}\label{static2}
  \langle t\rangle \equiv \int_{0}^{{1}} t P(t)dt =
  \int_{0}^{1} t \dot{\upsilon}(t)=
  \frac{1}{M^2}(2M-3+(M+3)e^{-M})
\end{equation}
This is in accord with the result of \cite{dm}. This formula
shows that the presence of a small number of shortcuts, causes a
significant drop in the average shortest path from $ 1$ to very
small values. In this sense the world gets
smaller by long range connections.\\
We now study the static effects of random directed bonds on a ring
without shortcuts. The presence of blocks makes the world small
in a different sense, namely for each site the number of
accessible sites gets smaller. In fact the average size of the
world accessible to a site is not $N$ anymore but it is given by
$ V({\frac{N}{2}})$ (see the paragraph leading to equation
(\ref{ddd})). Hence the probability of shortest paths is given by
$ P(T):= \frac{V(T)-V(T-1)}{V(\frac{N}{2})}$, or in the scaling
limit by
\begin{equation}\label{static4}
  P(t):= \frac{\dot{\upsilon}(t)}{\upsilon(\frac{1}{2})}
\end{equation}
This probability is normalized, i.e. $
\int_{0}^{\frac{1}{2}}P(t)dt = 1 $.  We obtain from
(\ref{static4})
\begin{equation}\label{static5}
\langle t\rangle =
\frac{1}{\upsilon(\frac{1}{2})}\int_{0}^{\frac{1}{2}} t
\dot{\upsilon}(t)dt
\end{equation}
However in order to assess the situation in this network, we
should compare the average shortest path with the size of this
small world itself, namely we should calculate $\frac {\langle t
\rangle }{\upsilon_0^{acc}}$. Inserting equation (\ref{dds}) into
(\ref{static5}) we find:
\begin{equation}\label{static6}
\frac{\langle t \rangle}{\upsilon_0^{acc}} = \frac{2-(\mu
+2)e^{\frac{-\mu}{2}}}{4(1-e^{-\frac{\mu}{2}})^2}
\end{equation}
Figure (\ref{fig-8}) shows both the average size of the
accessible world $\upsilon_0^{{acc}}$ and the ratio $\frac{\langle
t \rangle}{\upsilon_0^{acc}} $ of the average shortest path to
the size of accessible world as a function of the number of
blocks $ \mu $. It is seen that for $ \mu = 0 $, when there is no
block, the size is $ 1 $ and the average of the shortest path is
$\frac{1}{4}$ as it should be. With a few number of blocks the
size drops dramatically and the average of shortest path within
the world increases. Note that with increasing $ \mu $ the
average shortest path increases to its maximum value of $
\frac{1}{2}$.\\
For the fully random network, we use equations (\ref{eee}) and
(\ref{static4}) to obtain the average of shortest path. The result
is shown in figure (\ref{fig-9}) for several values of the
parameters.

\section{conclusion}
We have studied the effect of directed short and long range
connections in a simple model of small world network. In our
models all the shortcuts pass via a central site in the network.
This makes possible an almost exact calculation of many of the
properties of the network. We have calculated the function $V(T)$,
defined as the number of sites affected up to time $T$ when a
naive spreading process starts in the network. As opposed to
shortcuts, the presence of un-favorable bonds has a negative
effect on this quantity. Hence the spreading process may be able
to affect only a fraction of the total sites of the network. We
have defined this fraction to be the average size of the
accessible world in our model and have calculated it exactly for
our model. We have studied also the interplay of shortcuts, and
un-favorable bonds on the small world properties like the size of
accessible world, the speed of propagation of a spreading
process, and the average shortest path between two arbitrary
sites. Our results show that one can separately take into account
the effect of randomness in the directions of shortcuts and the
short-range connections in the underlying lattice and at the end
super-impose the two effects in a suitable way. We expect that
this will hold also in more complicated lattices of small world
networks.


\begin{thebibliography}{prsty}

 \bibitem{fhcs}L.F.L-Fernandez,R.Huerta,F.Corbacho and
 J.A.Siguenza,Phys. Rev. lett. {\bf 84}(12):2758-2761 Mar 2000.

 \bibitem{bm}
 J.W.Bohland and A.A.Minai,Neurocomputing {\bf 38-40}(2001),489-496.

\bibitem{jtao}
 H.Jeong,B.Tombor,R.Albert,Z.N.Oltvai and A.-L.Barabasi,Nature
{ \bf 407 }(2000),651-654.

 \bibitem{kas}
 R.V.Kulkarni,E.Almaas and D.Stroud,Phys. Rev. E {\bf 61}, 4268 (2000).

 \bibitem{plh}A.R.Puniyani,R.M.Lukose and B.A.
 Huberman,cond-mat/0107212.

 \bibitem{ajb}R. Albert, H. Jeong, and A. L. Barabasi, Nature,
 {\bf 401} :130-131 (1999).

 \bibitem{n2} M. E. J. Newman, Small Worlds,J. Stat. Phys. {\bf 101}(3-4):819-841 Nov 2000.

  \bibitem{deb}
 J.Davidsen,H.Ebel and S.Bornholdt, Phys. Rev. Lett. {\bf 88} 128701 (2002).

 \bibitem{ab1}
 R. Albert and A.-L.Barabasi, Rev. Mod. Phys. {\bf 74} 47 (2002).

 \bibitem{b} B. Bollaoas, {\it Random Graphs} Academic Press
 (New York)(1985).

 \bibitem{asbs}
 L.A.N.Amaral,A.Scala,M.Barthelemy and
 H.E.Stanley,Proc. Natl. Acad. Sci. USA {\bf 97}(2000):11149-11152.

\bibitem{ws}
 D.J.Watts and S.H.Strogatz , Nature {\bf 393},440(1998).

 \bibitem{bw}
 A.Barrat and M.Weigt.2000, Europ. Phys. J. B {\bf 13}, 547 (2000).

 \bibitem{nmw}
 M.E.J.Newman,C.Moore and D.J.Watts,Phys. Rev. Lett. {\bf 84},3201(2000).

 \bibitem{ka}  M.Kuperman and G.Abramson,Phys. Rev. Lett.  {\bf56}
 :1906 (2001).

 \bibitem{z}
 D.H.Zanette,Phys. Rev. E {\bf 64}(2001).

 \bibitem{jb}S.Jespersen and A.Blumen, Phys. Rev.
 E {\bf 62}(5):6270-6274 Part A ,Nov 2000.

 \bibitem{bkmr}
 A. Broder, R. Kumar, F. Maghoul, P. Raghavan, S. Rajagopalan,
  R. Stata, A. Tomikns and J. Wiener, Computer networks {\bf 33}(1-6), 2000, 309-320.

 \bibitem{t2}
 B.Tadic,Physica A {\bf 293}(2001),273-284.

 \bibitem{ste}
 O. Sporns,G. Tononi and G.M.Edelman,Neural Networks
 {\bf 13}(2000),909-922.

 \bibitem{slr}
 A.D.Sanchez,J.M.Lopez and M.A.Rodriguez, Phys. Rev. Lett. {\bf 88} 048701 (2002).

 \bibitem{t1}
 B.Tadic,cond-mat/0110033.

 \bibitem{kr}
 V.Karimipour and A.Ramzanpour, Phys. Rev. E {\bf 65} 36122 (2002).

 \bibitem{dm}
 S.N.Dorogovtsev and J.F.F.Mendes, Europhys. Lett. {\bf 50},1 (2000).

 \bibitem{aks}
 E.Almaas,R.V.Kullkarni and D.Stroud, Phys. Rev. Lett. {\bf 88}, 098101 (2002).

\end{thebibliography}
\end{document}